\documentclass[aps,prl,twocolumn,groupedaddress,preprintnumbers,nofootinbib]{revtex4}
\usepackage{graphicx}
\usepackage{amsfonts}
\usepackage{hyperref}
\usepackage{amsmath}
\usepackage{bm}
\begin{document}
\newcommand{\bra}{\langle}
\newcommand{\ket}{\rangle}
\newcommand{\al}{\alpha}
\newcommand{\be}{\beta}
\newcommand{\ga}{\gamma}
\newcommand{\de}{\delta}
\newcommand{\D}{\Delta}
\newcommand{\ep}{\epsilon}
\newcommand{\varep}{\varepsilon}
\newcommand{\e}{\eta}
\renewcommand{\th}{\theta}
\newcommand{\Th}{\Theta}
\newcommand{\la}{\lambda}
\newcommand{\La}{\Lambda}
\newcommand{\Ga}{\Gamma}
\newcommand{\m}{\mu}
\newcommand{\n}{\nu}
\renewcommand{\r}{\rho}
\newcommand{\si}{\sigma}
\newcommand{\Si}{\Sigma}
\newcommand{\ta}{\tau}
\newcommand{\vp}{\varphi}
\newcommand{\p}{\phi}
\renewcommand{\c}{\chi}
\newcommand{\ps}{\psi}
\renewcommand{\o}{\omega}
\renewcommand{\O}{\Omega}
\newcommand{\OO}{{\cal O}}
\newcommand{\C}{{\cal C}}
\newcommand{\pa}{\partial}
\newcommand{\beq}{\begin{equation}}
\newcommand{\eeq}{\end{equation}}
\newcommand{\mc}{\mathcal}
\newcommand{\mb}{\mathbb}
\newcommand{\mycomment}[1]{}
\newcommand{\wh}[1]{\widehat{#1}}
\newcommand{\fk}{f_\textup{bulk}}
\newcommand{\fy}{f_\textup{bdy}}
\newcommand{\yi}{y_i}
\newcommand{\yl}{y_L}
\newcommand{\yr}{y_R}

\newcommand{\nn}{\nonumber}
\newcommand{\Sl}{\sum\limits}
\newcommand{\blue}{\color{blue}}
\newcommand{\red}{\color{red}}
\newcommand{\green}{\color{green}}
\newcommand{\black}{\color{black}}

\def\msout#1{\textrm{\sout{#1}}}
\def\mxout#1{\textrm{\xout{#1}}}

 \def\be{\begin{equation}}
\def\ee{\end{equation}}
\def\bea{\begin{eqnarray}}
\def\eea{\end{eqnarray}}
\def\nn{\nonumber}

\def\tr{{\mbox{tr}}}
\def\Atr{{\mbox{Tr}}}

\def\a{\alpha}
\def\b{\beta}
\def\g{\gamma}
\def\d{\delta}
\def\lam{\lambda}
\def\u{\mu}
\def\v{\nu}
\def\r{\rho}
\def\t{\tau}
\def\z{\zeta}
\def\s{\sigma}
\def\th{\theta}

\def\te{\tilde{e}}
\def\tK{\tilde{K}}
\def\tB{\tilde{B}}
\def\htK{\hat{\tilde{K}}}

\def\Qh{\hat{Q}}
\def\baret{\overline{\eta}}
\def\homega{{\hat{\omega}}}
\def\bpsi{{\overline{\psi}}}
\def\wtau{{\widetilde{\tau}}}
\def\bth{{\overline{\theta}}}
\def\blam{{\overline{\lambda}}}

\def\da{{\dot{\a}}}
\def\db{{\dot{\b}}}
\def\dg{{\dot{\g}}}

\def\bj{{\overline{j}}}
\def\bk{{\overline{k}}}
\def\bz{{\overline{z}}}
\def\sa{{\hat{a}}}
\def\sb{{\hat{b}}}
\def\sc{{\hat{c}}}
\def\wa{{\tilde{a}}}
\def\wb{{\tilde{b}}}
\def\wc{{\tilde{c}}}
\def\oa{{\overline{a}}}
\def\ob{{\overline{b}}}
\def\oc{{\overline{c}}}
\def\od{{\overline{d}}}

 \def\CA{{\cal A}}
 \def\CC{{\cal C}}
 \def\CF{{\cal F}}
 \def\CI{{\cal I}}
 \def\cJ{{\cal J}}
 \def\tJ{{\tilde J}}
 \def\tcJ{{\tilde\cal J}} 
 \def\CO{{\cal O}}
 \def\o{{\rm ord}}
 \def\Ph{{\Phi }}
 \def\L{{\Lambda}}
 \def\CN{{\cal N}}
 \def\p{\partial}
 \def\pslash{\p \llap{/}}
 \def\Dslash{D \llap{/}}
 \def\apm{{\a^{\prime}}}
 \def\r{\rightarrow}
\def\ts{\tilde s}
\def\tu{\tilde u}
 \def\BR{\IR}
 \def\BZ{\IZ}
 \def\BC{\IC}
 \def\BM{\QM}
 \def\BP{\IP}
 \def\BH{\QH}
 \def\BX{\QX}
 \def\sym#1{{{\rm SYM}} _{#1 +1}}
 \def\imp{$\Rightarrow$}
 \def\IZ{\relax\ifmmode\mathchoice
 {\hbox{\cmss Z\kern-.4em Z}}{\hbox{\cmss Z\kern-.4em Z}}
 {\lower.9pt\hbox{\cmsss Z\kern-.4em Z}}
 {\lower1.2pt\hbox{\cmsss Z\kern-.4em Z}}\else{\cmss Z\kern-.4em Z}\fi}
 \def\IB{\relax{\rm I\kern-.18em B}}
 \def\IC{{\relax\hbox{$\inbar\kern-.3em{\rm C}$}}}
 \def\Ic{{\relax\hbox{$\inbar\kern-.22em{\rm c}$}}}
 \def\ID{\relax{\rm I\kern-.18em D}}
 \def\IE{\relax{\rm I\kern-.18em E}}
 \def\IF{\relax{\rm I\kern-.18em F}}
 \def\IG{\relax\hbox{$\inbar\kern-.3em{\rm G}$}}
 \def\IGa{\relax\hbox{${\rm I}\kern-.18em\Gamma$}}
 \def\IH{\relax{\rm I\kern-.18em H}}
 \def\II{\relax{\rm I\kern-.18em I}}
 \def\IK{\relax{\rm I\kern-.18em K}}
 \def\IP{\relax{\rm I\kern-.18em P}}
 \def \glc {\text{GL}(1, \mathbb{C})}

\def\Tr{{\rm Tr}}
 \font\cmss=cmss10 \font\cmsss=cmss10 at 7pt
 \def\IR{\relax{\rm I\kern-.18em R}}

\def\wdg{{\wedge}}

\newcommand\ev[1]{{\langle {#1}\rangle}}
\newcommand\SUSY[1]{{{\cal N} = {#1}}}
\newcommand\diag[1]{{\mbox{diag}({#1})}}
\newcommand\com[2]{{\left\lbrack {#1}, {#2}\right\rbrack}}

\newcommand\px[1]{{\partial_{#1}}}
\newcommand\qx[1]{{\partial^{#1}}}

\newcommand\rep[1]{{\bf {#1}}}

\def\gam{{\widetilde{\gamma}}} 
\def\sig{{\sigma}} 
\def\hsig{{\hat{\sigma}}} 

\def\eps{{\epsilon}} 

\def\bZ{{\overline{Z}}}
\def\BR{{\mathbb R}}
\def\Lag{{\cal L}}
\def\cO{{\cal O}}
\def\cH{{\cal H}}
\def\wcO{{\widetilde{\cal O}}}
\def\vL{{\vec{L}}}
\def\vx{{\vec{x}}}
\def\vy{{\vec{y}}}

\def\vLf{{\vec{\lambda}}}

\newcommand\cvL[1]{{L^{#1}}} 

\def\npsi{{\psi^{(inv)}}}
\def\bnpsi{{\bpsi^{(inv)}}}
\def\wpsi{{\widetilde{\psi}}}
\def\bwpsi{{\overline{\widetilde{\psi}}}}
\def\hpsi{{\hat{\psi}}}
\def\bhpsi{{\overline{\hat{\psi}}}}

\newcommand\nDF[1]{{{D^F}_{#1}}}
\def\wA{{\widetilde{A}}}
\def\wF{{\widetilde{F}}}
\def\wJ{{\widetilde{J}}}
\def\hA{{\hat{A}}}
\def\hF{{\hat{F}}}
\def\hJ{{\hat{J}}}
\def\MapL{{\Upsilon}}
\def\hR{{\hat{R}}}
\def\hL{{\hat{L}}}

\def\cpl{{\lambda}}  
\def\mcr{{\mathcal{R}}}
\def\xv{{\vec{x}}}
\def\xvt{{\vec{x}^{\top}}}
\def\nv{{\hat{n}}}
\def\nvt{{\hat{n}^{\top}}}
\def\hM{{M}}
\def\hgM{{\widetilde{M}}}
\def\utr{{\mbox{tr}}}

\newcommand{\fig}[1]{fig.\ (\ref{#1})}
\def\dd{\mbox{d}}
\def\ddd{\mbox{\sm d}}
\def\o{\omega}
\def\bra{\langle}
\def\ket{\rangle}
\def\a{\alpha}
\def\b{\beta}
\def\d{\delta}
\def\dd{\partial}
\def\D{\Delta}
\def\LL{\triangle}
\def\g{\gamma}
\def\G{\Gamma}
\def\e{\epsilon}
\def\ve{\varepsilon}
\def\et{\eta}
\def\f{\phi}
\def\F{\Phi}
\def\vf{\varphi}
\def\k{\kappa}
\def\l{\lambda}
\def\L{\Lambda}
\def\m{\mu}
\def\n{\nu}
\def\s{\sigma}
\def\S{\Sigma}
\def\o{\omega}
\def\p{\pi}
\def\r{\rho}
\def\t{\tau}
\def\th{\theta}
\def\vt{\vartheta}
\def\ra{\rightarrow}
\def\la{\leftarrow}
\def\pa{\partial}
\def\ov{\overline}
\def\Pl{s_{\sm{Pl}}}
\def\tr{\tilde R}
\def\td{\tilde d}
\def\gmn{g_{\mu \nu}}
\def\DO{\D_2}
\def\O{{\cal O}}
\def \slc {{SL}(2, \mathbb{C})}
\def \vol {\text{vol}\,}

\newcommand{\ti}[1]{\tilde{#1}}
\renewcommand{\^}[1]{\hat{#1}}
\newcommand{\sm}[1]{\mbox{\scriptsize #1}}
\newcommand{\tn}[1]{\mbox{\tiny #1}}
\renewcommand{\@}[1]{\sqrt{#1}}
\renewcommand{\le}[1]{\label{#1}\end{eqnarray}}
\newcommand{\eq}[1]{(\ref{#1})}
\def\nn{\nonumber\\}
\def\nm{\nonumber}
\newcommand{\rf}[1]{\cite{ref:#1}}
\newcommand{\rr}[1]{\bibitem{ref:#1}}
\def\qu{\ {\buildrel{\displaystyle ?} \over =}\ }
\def\smqu{\ {\buildrel ?\over =}\ }
\def\ffract#1#2{\raise .35 em\hbox{$\scriptstyle#1$}\kern-.25em/
\kern-.2em\lower .22 em \hbox{$\scriptstyle#2$}}
\def\GN{G_{\mbox{\tn N}}}
\def\lPl{s_{\mbox{\tn{Pl}}}}
\def\fn{f_{(0)}}
\def\fe{f_{(1)}}
\def\ft{f_{(2)}}
\def\fd{f_{(3)}}
\def\Ric{{\mbox{Ric}}}
\def\Rie{{\mbox{Rie}}}
\def\Ein{{\mbox{Ein}}}
\def\nl{\newline}
\def\cl{\textcolor}
\def\clb{\colorbox}
\def\na{\nabla}
\def\half{{1\over2}\,}
\def\nonu{\nonumber \\{}}
\def\da{\dot{a}}
\def\db{\dot{b}}
\def\dc{\dot{c}}
\def\tg{\tilde{\g}}
\def\bF{\bar{F}}


\def\vwsX{{\bf X}} 
\def\dvwsX{{\dot{\bf X}}} 
\newcommand\vxmod[1]{{{\bf X}_{#1}}} 
\newcommand\dvxmod[1]{{\dot{{\bf X}}_{#1}}} 

\def\vwsP{{\bf \Pi}} 

\def\wsX{{X}} 
\def\wsZ{{Z}} 
\def\bwsZ{{\overline{Z}}} 
\def\wsPsi{{\Psi}} 
\def\bwsPsi{{\overline{\Psi}}} 

\def\twsZ{{\widetilde{Z}}} 
\def\twS{{\widetilde{S}}} 
\def\btwsZ{{\overline{\widetilde{Z}}}} 
\def\twsPsi{{\widetilde{\Psi}}} 
\def\btwsPsi{{\overline{\widetilde{\Psi}}}} 
\def\tV{{\widetilde{V}}}
\def\talp{{{\tilde{\a}}'}} 

\def\hLf{{\hat{\lambda}}}
\def\hL{{\hat{L}}}

\newdimen\tableauside\tableauside=1.0ex
\newdimen\tableaurule\tableaurule=0.4pt
\newdimen\tableaustep
\def\phantomhrule#1{\hbox{\vbox to0pt{\hrule height\tableaurule width#1\vss}}}
\def\phantomvrule#1{\vbox{\hbox to0pt{\vrule width\tableaurule height#1\hss}}}
\def\sqr{\vbox{%
  \phantomhrule\tableaustep
  \hbox{\phantomvrule\tableaustep\kern\tableaustep\phantomvrule\tableaustep}%
  \hbox{\vbox{\phantomhrule\tableauside}\kern-\tableaurule}}}
\def\squares#1{\hbox{\count0=#1\noindent\loop\sqr
  \advance\count0 by-1 \ifnum\count0>0\repeat}}
\def\tableau#1{\vcenter{\offinterlineskip
  \tableaustep=\tableauside\advance\tableaustep by-\tableaurule
  \kern\normallineskip\hbox
    {\kern\normallineskip\vbox
      {\gettableau#1 0 }%
     \kern\normallineskip\kern\tableaurule}%
  \kern\normallineskip\kern\tableaurule}}
\def\gettableau#1 {\ifnum#1=0\let\next=\null\else
  \squares{#1}\let\next=\gettableau\fi\next}

\tableauside=1.0ex
\tableaurule=0.4pt



\newcommand{\eeqn}{\end{eqnarray}}
\newcommand{\ack}[1]{{\bf Pfft! #1}}
\newcommand{\osigma}{\overline{\sigma}}
\newcommand{\orho}{\overline{\rho}}
\newcommand{\myfig}[3]{
	\begin{figure}[ht]
	\centering
	\includegraphics[width=#2cm]{#1}\caption{#3}\label{fig:#1}
	\end{figure}
	}
\newcommand{\littlefig}[2]{
	\includegraphics[width=#2cm]{#1}}

\preprint{QMUL-PH-18-30, \, CALT-TH-2018-054}
\title{\bf{All Tree Amplitudes of 6D $\mathbf{(2,0)}$ Supergravity:\\ Interacting Tensor Multiplets and the $\mathbf{\mathrm{K}}3$ Moduli Space}}
\author{{\bf Matthew Heydeman$^G$, John H. Schwarz$^{G}$, Congkao Wen$^{T}$, and Shun-Qing Zhang$^{T}$ }}

\affiliation{$^G$ Walter Burke Institute for Theoretical Physics, California Institute of Technology, Pasadena, CA  91125, USA.\\
  $^{T}$ Centre for Research in String Theory,
School of Physics \& Astronomy, \\ Queen Mary University of London, UK.
}

\begin{abstract}
We present a twistor-like formula for the complete tree-level S matrix of 6D $(2,0)$ supergravity coupled to $21$ abelian tensor multiplets. This is the low-energy effective theory that corresponds to Type IIB superstring theory compactified on a $\mathbf{\mathrm{K}}3$ surface. The formula is expressed as an integral over the moduli space of certain rational maps of the punctured Riemann sphere. By studying soft limits of the formula, we are able to explore the local moduli space of this theory, ${SO(5,21)\over SO(5)\times SO(21)}$. Finally, by dimensional reduction, we also obtain a new formula for the tree-level S matrix of 4D $\mathcal{N}=4$ Einstein-Maxwell theory.  
\end{abstract}

\maketitle

\section{Introduction}

To describe scattering amplitudes of supersymmetric theories in higher dimensions, \cite{Heydeman:2017yww, Cachazo:2018hqa} introduced a six-dimensional rational map formalism in the spirit of~\cite{Witten:2003nn,Roiban:2004yf,Cachazo:2013iaa}. Using this formalism, extremely compact formulas were found for tree-level amplitudes of a wide range of interesting theories, including maximally supersymmetric gauge theories and supergravity in diverse dimensions, as well as the world-volume theories of probe D-branes and the M5-brane in flat space. In the case of the M5-brane~\cite{Heydeman:2017yww}, which contains a chiral tensor field, the formalism circumvents a common difficulty in formulating a covariant action principle due to the self-duality constraint.

In this article, we continue to explore the utility of the 6D rational maps and spinor-helicity formalism and present the tree-level S matrix for the theory of 6D $(2,0)$ supergravity. This chiral theory arises as the low-energy limit of Type IIB string theory compactified on a $\mathrm{K}3$ surface~\cite{Witten:1995zh} and is particularly interesting because it describes the interaction of self-dual tensors and gravitons. 

To describe massless scattering in 6D, it is convenient to introduce spinor-helicity variables~\cite{Cheung:2009dc},
\bea
p_i^{AB} = \lambda^A_{i, a} \lambda^{B}_{i, b} \epsilon^{ab} := \langle \lambda^A_{i}\, \lambda^{B}_i \rangle. 
\eea
Here, and throughout, $i = 1,\dots, n$ labels the $n$ particles, $A=1, 2, 3, 4$ is a spinor index of the $Spin(5,1)$ Lorentz group, and $a = 1,2$ is a left-handed index of the $SU(2)_L \times SU(2)_R$ massless little group. This is the only non-trivial little-group information that enters for chiral $(2,0)$ supersymmetry--the $(2,0)$ supergravity multiplet and a number of $(2,0)$ tensor multiplets, which contain a chiral tensor. The tensor multiplets transform as singlets of $SU(2)_R$, whereas the gravity multiplet is a triplet; later we will introduce the doublet index $\hat{a}$ for $SU(2)_R$. 

We also introduce a flavor index $f_i$ with $i= 1, \dots, 21$ to label the 21 tensor multiplets; this is the number that arises in 6D from compactification of the NS and R fields of Type IIB superstring theory on a $\mathrm{K}3$ surface. It is also the unique number for which the gravitational anomalies cancel~\cite{comment-anomaly}. We assume that we are at generic points of the moduli space, where perturbative amplitudes are well-defined~\cite{comment-moduli}. Interestingly one can explore the moduli space of the theory from the S matrix by studying soft limits~\cite{ArkaniHamed:2008gz}. Indeed, we derive new soft theorems from the formula we construct, which describe precisely the moduli space of 6D $(2,0)$ supergravity: ${SO(5, 21) \over SO(5)\times SO(21)}$.

In the rational-map formulation, amplitudes for $n$ particles are expressed as integrals over the moduli space of rational maps from the $n$-punctured Riemann sphere to the space of spinor-helicity variables. In general, the amplitudes take the following form~\cite{Cachazo:2013hca, Heydeman:2017yww, Cachazo:2018hqa}, 
\bea
A^{6D}_n = \int d\mu^{\rm 6D}_n \, \mathcal{I}_L \, \mathcal{I}_R \, ,
\eea
where $d\mu^{\rm 6D}_n$ is the measure encoding the 6D kinematics and the product $\mathcal{I}_L \, \mathcal{I}_R$ is the integrand that contains the dynamical information of the theories, including supersymmetry. The measure is given by
\bea
d\mu^{\rm 6D}_n = \frac{\prod_{i=1}^n d\sigma_i\, \prod_{k=0}^{m} d^8 \rho_k}{\vol( \slc_\sigma \times \slc_\rho)} \frac{1}{V_n^2}\prod_{i=1}^n E^{\rm 6D}_i ,
\eea
and $n=2m+2$ (we will discuss $n=2m+1$ later). The coordinates $\s_i$ label the $n$ punctures, and $V_n = \prod_{i<j} \s_{ij}$, with $\sigma_{ij} =\s_i - \s_j$. They are determined up to an overall $\slc_\sigma$ M{\"o}bius group transformation, whose ``volume" is divided out in a standard way. The 6D scattering equations are given by
\bea \label{eq:6DSE}
E^{\rm 6D}_i = \delta^6 \left( p^{AB}_i - \frac{\langle \rho^{A}(\sigma_i)\, \rho^{B}(\sigma_i) \rangle }{\prod_{j\neq i} \sigma_{ij}}\right).
\eea
These maps are given by degree-$m$ polynomials 
$
\rho^{A}_a(\sigma) = \sum_{k=0}^{m} \rho^A_{a,k}\, \sigma^k ,
$
which are determined up to an overall $\slc_\rho$ transformation, whose volume is divided out. This group is a complexification of $SU(2)_L$. 

It is straightforward to see that (\ref{eq:6DSE}) implies the on-shell conditions $p_i^2=0$ and momentum conservation. Furthermore as shown in~\cite{Heydeman:2017yww, Cachazo:2018hqa}, this construction implies that the integrals are completely localized on the $(n-3)!$ solutions, which are equivalent to those of the general dimensional scattering equations~\cite{Cachazo:2013hca}, 
\begin{align}
\sum_{i\neq j} {p_i \cdot p_j \over \sigma_{ij}} =0\,, \quad {\rm for ~~all} ~~j \, .
\end{align}
As we will see shortly that, unlike the general-dimensional scattering equations, the use of the spinor-helicity coordinates and 6D scattering equations allows us to make supersymmetry manifest.

Now consider $n=2m+1$, for which we have~\cite{Cachazo:2018hqa},
\begin{equation}
\!\! d\mu_{n}^{\text{6D}} = \frac{\left(\prod_{i=1}^{n} d\sigma_i \prod_{k=0}^{m-1}d^{8}\rho_{k}\right)\, d^{4}\omega\, \langle \xi d\xi\rangle }{\vol \left( \slc_{\sigma} , \slc_{\rho} , {T} \right)}\, \frac{1}{V^{2}_n} \prod_{i=1}^{n} E_i^{\rm 6D}. \label{eq:prop}
\end{equation}
The polynomials $\rho^{A}_a(\sigma)$ now are given by
\bea \label{eq:odd-pt-map}
\rho^{A}_a(\sigma) = \sum_{k=0}^{m-1} \rho^A_{a,k}\, \sigma^k + \omega^A \xi_a\, \sigma^{m},
\eea
and there is a shift symmetry ${T}(\a)$ acting on $\omega^A$: $\omega^A \rightarrow \omega^A + \alpha \langle \xi \, \rho^A_{m-1}\rangle$, which we also have to mod out.

Here we review the integrand factors for 6D $(2,2)$ supergravity, since they will be relevant. For $(2,2)$ supergravity, we have,
\bea \label{eq:(2,2)-integrand}
 \mathcal{I}_L = { {\rm det}^{\prime} S_n } , \quad \mathcal{I}_R =  \Omega_F^{(2,2)} ,
\eea
where $S_n$ is a $n\times n$ anti-symmetric matrix, with entries: $\left[S_{n}\right]_{ij} = {p_i \cdot p_j \over \sigma_{ij} }$. This matrix has rank $(n{-}2)$, and  the reduced Pfaffian and determinant are defined as
\bea
{\rm Pf}^{\prime} S_n = {(-1)^{i+j} \over \sigma_{ij} } {\rm Pf}S^{ij}_{ij}, \quad 
{\rm det}^{\prime} S_n = \left({\rm Pf}^{\prime} S_n  \right)^2 . 
\eea
Here $S^{ij}_{ij}$ means that the $i$-th and $j$-th rows and columns of $S_n$ are removed, and the result is $i, j$ independent~\cite{Cachazo:2014xea}. $\Omega_F^{(2,2)}$ is a fermionic function of Grassmann coordinates $\eta_i^{Ia}, \tilde{\eta}_i^{\hat{I} \hat{a}}$, which we use to package the supermultiplet of on-shell states into a `superfield',
\begin{align}  \label{eq:(2,2)field}
\Phi^{(2,2)}&(\eta, \tilde{\eta})
=  \phi'  + \cdots + \eta^I_a \eta_{I,b}B^{ab} + \tilde{\eta}^{\hat I, \hat a} \tilde{\eta}^{\hat b}_{\hat I}B_{{\hat a} {\hat b}}  \\
&+\cdots+ \eta^I_a {\eta}_{I,b} \tilde{\eta}^{\hat I}_{\hat a} \tilde{\eta}_{\hat {I}, \hat {b} } G^{ab \hat{a} \hat{b}}  
 +\, \cdots + (\eta)^4 (\tilde{\eta})^4 \bar{\phi'} , \nonumber 
\end{align}
where $B^{ab}$ and $B_{{\hat a} {\hat b}}$ are self-dual and anti self-dual two forms, and $G^{ab \hat{a} \hat{b}}$ is the graviton. Here $I, \hat{I}=1,2$ are the R-symmetry indices corresponding to a $SU(2)\times SU(2)$ subgroup of the full $USp(4) \times USp(4)$ R-symmetry. The fermionic function $\Omega_F^{(2,2)}$ imposes the conservation of supercharge, which may be viewed as a double copy: $\Omega_F^{(2,2)} = \Omega_F^{(2,0)} \, \Omega_F^{(0,2)},
$
and $\Omega_F^{(2,0)}$ is given by
\bea \label{eq:(2,0)}
\Omega_F^{(2,0)} =V_n \prod_{k=0}^{m} \delta^4 \left(\sum_{i=1}^n {C}_{a,k;i, b} \, {\eta}^{I b}_i \right).
\eea
The $n\times 2n$ matrices ${C}_{a,k;i, b}= ({W}_i)_{a}^{b} \sigma_i^k$ and $({W}_i)_{a}^{b}$ can be expressed in terms of $\rho^{A}_a(\sigma_i)$ via
\bea 
p_i^{AB} W^a_{i,b} =\frac{\rho^{[A,a}(\sigma_i) \, \lambda^{B]}_{i,b}}{\prod_{j\neq i}\sigma_{ij}},
\eea
which is independent of $A, B$, and satisfies $\det W_{i}=\prod_{j\neq i}\sigma_{ij}^{-1}$. The matrix ${C}_{a,k;i, b}$ is a symplectic Grassmannian which was used in~\cite{Cachazo:2018hqa} as an alternative way to impose the 6D scattering equations. $\Omega_F^{(0,2)}$ is the conjugate of $\Omega_F^{(2,0)}$, and the definition is identical, with the understanding that we use the right-handed variables, such as $\tilde{\eta}^{\hat{I}}_{\hat{a}}, \tilde{\lambda}_{\hat{A} \hat{a}}, \tilde{\rho}_{\hat{A} \hat{a}},\tilde{\xi}_{\hat a}, (\tilde{W}_i)_{\hat a}^{\hat b}$, etc.

For $n=2m+1$, the integrands take a slightly different form. For the fermionic part, we have
\bea \label{eq:(2,0)odd}
\Omega_F^{(2,0)} &=V_n \prod_{k=0}^{m-1} \delta^4 \left(\sum_{i=1}^n {C}_{a,k;i, b} \, {\eta}^{I b}_i \right) \\ \nonumber
&\times\, \delta^2\left(\sum_{i=1}^n \xi^a {C}_{a,m;i, b} \, {\eta}^{I b}_i \right) . 
\eea 
whereas the $n\times n$ matrix $S_n$ is modified to an $(n+1)\times (n+1)$ matrix, which we denote $\hat{S}$. $\hat{S}_n$ is defined in the same way as ${S}_n$, but with $i,j=1,\ldots,n,\star$. Here $\sigma_\star$ is a reference puncture, and $p_{\star}$ is given by
\be \label{eq:pstar}
p_{\star}^{AB} = \dfrac{ 2\, q^{[A} p^{B]C}(\sigma_\star) \tilde{q}_C }{ q^D  [\tilde{\rho}_D(\sigma_\star )\, \tilde{\xi}] \langle \rho^E(\sigma_\star)\, \xi \rangle  \tilde{q}_E},
\ee
where $q$ and $\tilde{q}$ are arbitrary spinors. 

\vspace{-0.15cm}

\section{6D $(2,0)$ Supergravity}
The 6D $(2,0)$ supergravity theory contains $21$ tensor multiplets and the graviton multiplet. The superfield of the tensor multiplet is a singlet of the little group,
\begin{eqnarray}
\Phi(\eta)= \phi + \cdots + \eta^{I}_a \eta_{I,b} B^{ab} + \cdots  + (\eta)^4 \bar{\phi} , 
\end{eqnarray}
where $a, b=1,2$ are the $SU(2)_L$ little-group indices.  The graviton multiplet transforms as a $(\mathbf{1}, \mathbf{3})$ of the little group, so the superfield carries explicit $SU(2)_R$ indices,
\bea
\Phi_{\hat{a} \hat{b}}(\eta)
=  B_{\hat{a} \hat{b}} + \ldots+ \eta^{I}_a \eta_{I,b} G^{ab}_{\hat{a} \hat{b}} + \ldots  + (\eta)^4 \bar{B}_{\hat{a} \hat{b}} \, ,
\eea
and $\Phi_{\hat{a} \hat{b}}(\eta) = \Phi_{\hat{b} \hat{a}}(\eta)$.
We see that both the tensor multiplet and graviton multiplet can be obtained from the 6D $(2,2)$ superfield in (\ref{eq:(2,2)field}) via SUSY reductions~\cite{Elvang:2011fx}\cite{comment-susyred}, 
\bea \label{eq:SUSY_Red1}
\Phi(\eta) &=&\int d \tilde{\eta}^{\, \hat{I}}_{\hat{a}}  d \tilde{\eta}_{\, \hat{I}}^{\hat{a}}\, \Phi^{(2,2)}(\eta, \tilde{\eta}) \big{|}_{ \tilde{\eta} \rightarrow 0} \, ,
\cr
\Phi_{\hat{a} \hat{b}}(\eta) &=& \int d \tilde{\eta}^{\, \hat{I}}_{\hat{a}} d \tilde{\eta}_{{\hat{I}} \hat{b}} \, \Phi^{(2,2)}(\eta, \tilde{\eta}) \big{|}_{ \tilde{\eta} \rightarrow 0} \, .
\eea
These integrals have the effect of projecting onto the right-handed $USp(4)$ R-symmetry singlet sector, which reduces $(2,2) \rightarrow (2,0)$. Using the reduction, the amplitudes of $(2,0)$ supergravity with $n_1$ supergravity multiplets and $n_2$ tensor multiplets of the same flavor ($n_1 + n_2 =n$) can be obtained from  the $(2,2)$ supergravity amplitude via
\bea
 A^{(2,0)}_{n_1, n_2} =\!\! \int\! \prod_{i \in n_1} d\tilde{\eta}^{\hat{I}}_{i, \hat{a}_i } d \tilde{\eta}_{i, {\hat{I}} \hat{b}_i} \prod_{j \in n_2} d\tilde{\eta}^{\hat{J}}_{j, \hat{a}_j } d \tilde{\eta}^{ \, \hat{a}_j}_{j, \hat{J}}    A^{(2,2)}_n (\eta, \tilde{\eta}). \nonumber
\eea
Note $A^{(2,2)}_n (\eta, \tilde{\eta}) \sim \eta^{2n} \tilde{\eta}^{2n}$, so the integration removes all $\tilde{\eta}$'s. The fermionic integral can be performed using (\ref{eq:(2,2)-integrand}), and (\ref{eq:(2,0)}) (or (\ref{eq:(2,0)odd}) for odd $n$), and we obtain
\bea
A^{(2,0)}_{n_1, n_2}  =    \int d \mu_n^{\rm 6D} \tilde{M}_{\hat{a} \hat{b}}^{n_1 n_2}\, V_n\,  {\rm det}^{\prime} S_n  \, \Omega_F^{(2,0)} ,
\eea
where $\tilde{M}_{\hat{a} \hat{b}}^{n_1 n_2}$, which we will define shortly, is obtained by integrating out $\Omega_F^{(0,2)}$. 

We begin with $n$ even, as the odd-$n$ case works in a similar fashion. Introducing the $n\times n$ matrix
\bea
\!\!\!\!\! \tilde{M}_{\hat{a}_1 \cdots \hat{a}_n} = \left( \begin{array}{cccc}
\tilde{C}_{\hat 1,0;1, \hat a_1}  &\tilde{C}_{\hat 1,0;2, \hat a_2} & \cdots & \tilde{C}_{\hat 1,0;n, \hat a_n} \\
\vdots  & \vdots  & \cdots & \vdots  \\
\tilde{C}_{\hat 1,m;1, \hat a_1}  &\tilde{C}_{\hat 1,m;2, \hat a_2} & \cdots & \tilde{C}_{\hat 1,m;n, \hat a_n}  \\
\tilde{C}_{\hat 2,0;1, \hat a_1}  &\tilde{C}_{\hat 2,0;2, \hat a_2} & \cdots & \tilde{C}_{\hat 2,0;n, \hat a_n} \\
\vdots  & \vdots  & \cdots & \vdots 
 \\
\tilde{C}_{\hat 2,m;1, \hat a_1}  &\tilde{C}_{\hat 2,m;2, \hat a_2} & \cdots & \tilde{C}_{\hat 2,m;n, \hat a_n} 
\end{array}  \right),
\eea
then $\tilde{M}_{\hat{a} \hat{b}}^{n_1 n_2}$  is given by 
\bea \label{eq:MM}
\tilde{M}_{\hat{a} \hat{b}}^{n_1 n_2}= \det \tilde{M}_{\hat{a}_1 \cdots \hat{a}_n} \det \tilde{M}_{\hat{b}_1 \cdots \hat{b}_n}  .
\eea
Note that here $\hat{a}$ and $\hat{b}$ denote sets of indices.
The indices $\hat{a}_i, \hat{b}_i$ are contracted if $i \in n_2$, whereas for $j \in n_1$ we symmetrize $\hat{a}_j, \hat{b}_j$. This corresponds to constructing little-group singlets for tensor multiplets and triplets for graviton multiplets. After the contraction and symmetrization, the result of (\ref{eq:MM}) simplifies drastically~\cite{comment-simplification} 
\bea \label{eq:single-flavor}
\tilde{M}_{\hat{a} \hat{b}}^{n_1 n_2} \rightarrow { {\rm Pf}  {X}_{n_2}  \over V_{n_2} }\, \tilde{M}_{\hat{a} \hat{b}}^{n_1 0} \,,
\eea
where $X_{n_2}$ is a $n_2 \times n_2$ anti-symmetric matrix given by
\bea
\left[X_{n_2}\right]_{ij} =\begin{cases}
               {1 \over \sigma_{ij} } \quad ~ \! {\rm if}  \quad i\neq j \\
            0  \quad \quad {\rm if}  \quad i = j\, ,
            \end{cases}
\eea
and $\tilde{M}_{\hat{a} \hat{b}}^{n_1 0}$ contains only the graviton multiplets. Let's remark that the simplification (\ref{eq:single-flavor}) (especially the appearance of ${\rm Pf}X_{n_2}$) will be crucial for the generalization to amplitudes with multiple tensor flavors which is more interesting and relevant for type IIB on $\mathbf{\mathrm{K}}3$. 

At this point in the analysis, we have obtained the tree-level amplitudes of 6D $(2,0)$ supergravity with a single flavor of tensor multiplets:
\bea
\!\!\!\!\!\! A^{(2,0)}_{n_1, n_2} =    \int d \mu_n^{\rm 6D} { {\rm Pf}  {X}_{n_2}  \over V_{n_2} } \, \tilde{M}_{\hat{a} \hat{b}}^{n_1 0} \, V_n\,  {\rm det}^{\prime} S_n \, \Omega_F^{(2,0)}.
\eea
The factor ${\rm Pf} {X}_{n_2}$ requires the non-vanishing amplitudes to contain an even number $n_2$ of tensor multiplets, as expected. For odd $n$, the matrix $\tilde{M}_{\hat{a}_1 \cdots \hat{a}_n}$ is given by
\begin{align} \label{eq:odd_n-M}
\!\!\!\!\!\!\!\!\! {\tilde{M}}_{\hat{a}_1 \cdots \hat{a}_n} = \left( \begin{array}{cccc}
\tilde{\xi}^{\hat b} \tilde{C}_{\hat b,m;1, \hat a_1}  & \tilde{\xi}^{\hat b} \tilde{C}_{\hat b,m;2, \hat a_2} & \cdots & \tilde{\xi}^{\hat b} \tilde{C}_{\hat b,m;n, \hat a_n} \\
\tilde{C}_{\hat 1,0;1, \hat a_1}  &\tilde{C}_{\hat 1,0;2, \hat a_2} & \cdots & \tilde{C}_{\hat 1,0;n, \hat a_n} \\
\vdots  & \vdots  & \cdots & \vdots  \\
\tilde{C}_{\hat 1,m-1;1, \hat a_1}  &\tilde{C}_{\hat 1,m-1;2, \hat a_2} & \cdots & \tilde{C}_{\hat 1,m-1;n, \hat a_n}  \\
\tilde{C}_{\hat 2,0;1, \hat a_1}  &\tilde{C}_{\hat 2,0;2, \hat a_2} & \cdots & \tilde{C}_{\hat 2,0;n, \hat a_n} \\
\vdots  & \vdots  & \cdots & \vdots 
 \\
\tilde{C}_{\hat 2,m-1;1, \hat a_1}  &\tilde{C}_{\hat 2,m-1;2, \hat a_2} & \cdots & \tilde{C}_{\hat 2,m-1;n, \hat a_n} ,
\end{array}  \right) \nonumber
\end{align}
recall $\tilde\xi^{\hat{b}}$ is the right-hand version of $\xi^{{b}}$ in (\ref{eq:odd-pt-map}).
Then the amplitudes take the same form
\bea
\!\! A^{(2,0)}_{n_1, n_2}  =    \int d \mu_n^{\rm 6D} { {\rm Pf}  {X}_{n_2}  \over V_{n_2} } \, {\tilde{M}}_{\hat{a} \hat{b}}^{n_1 0}\,V_n\,  {\rm det}^{\prime} \hat{S}_n \, \Omega_F^{(2,0)} .
\eea

\subsection{Multi-flavor tensor multiplets}
As we have emphasized, the identity (\ref{eq:single-flavor}) is crucial for the generalization to multiple tensor flavors, which is required for the 6D $(2,0)$ supergravity. Indeed, the formula takes a form similar to that of a Einstein-Maxwell theory worked out by Cachazo, He and Yuan~\cite{Cachazo:2014xea}, especially the object ${\rm Pf}X_{n_2}$. In that case, in passing from single-$U(1)$ photons to multiple-$U(1)$ ones, one simply replaced the matrix $X_n$ by $\mathcal{X}_n$~\cite{Cachazo:2014xea}, 
\bea \label{eq:sn-even}
\left[\mathcal{X}_{n}\right]_{ij} =\begin{cases}
               {\delta_{f_i f_j} \over \sigma_{ij} } \quad ~ \! {\rm if}  \quad i\neq j \\
            0  \quad \quad ~~ {\rm if}  \quad i = j
            \end{cases}
\eea
 which allows the introduction of multiple distinct flavors: namely, $f_i, f_j$ are flavor indices, and $\delta_{f_i f_j}=1$ if particles $i, j$ are of the same flavor, otherwise $\delta_{f_i f_j}=0$. Inspired by this result, we are led to a proposal for the complete tree-level S matrix of 6D $(2,0)$ supergravity with multiple flavors of tensor multiplets:
\bea \label{eq:multi-tensor}
\!\! \boxed{A^{(2,0)}_{n_1, n_2} = \! \int d \mu_n^{\rm 6D} \, { {\rm Pf}  {\mathcal{X}}_{n_2}  \over V_{n_2} } \, \tilde{M}_{\hat{a} \hat{b}}^{n_1 0} \, V_n\,  {\rm det}^{\prime} S_n \, \Omega_F^{(2,0)}.}
\eea
Again, the 6D scattering equations and integrands take different forms depending on whether $n$ is even or odd~\cite{comment:even-odd}. Since $n_2$ is necessarily even, this is equivalent to distinguishing whether $n_1$ is even or odd. 

Equation (\ref{eq:multi-tensor}) is our main result, which is a localized integral formula that describes all tree-level superamplitudes of abelian tensor multiplets (with multiple flavors) coupled to gravity multiplets. We can verify that it has all the correct properties. For instance, due to the fact that all the building blocks of the formula come from either 6D $(2,2)$ supergravity or Einstein-Maxwell theory, they all behave properly in the factorization limits, and transform correctly under the symmetries: $\slc_\sigma$, $\slc_\rho$, etc. Also, as we will show later, when reduced to 4D the proposed formula produces (supersymmetric) Einstein-Maxwell amplitudes, which is another consistency check. Finally, it is straightforward to check that the formula gives correct low-point amplitudes, e.g.~\cite{Lin:2015dsa}
\bea \label{eq:4pts}
\!\!\! A^{(2,0)}_{0,4}  &=& \delta^8 (Q) \left( { \delta^{f_1 f_2} \delta^{f_3 f_4} \over s_{12} }  + { \delta^{f_1 f_3} \delta^{f_2 f_4} \over s_{13} } +  { \delta^{f_2 f_3} \delta^{f_1 f_4} \over s_{23} }  \right), \cr
A^{(2,0)}_{2,2} &=& \delta^{f_1 f_2} { \delta^8 (Q)  [1_{\hat{a}_1} 2_{ \hat{a}_2} 3_{ \hat{a}_3} 4_{ \hat{a}_4}] [1^{\hat{a}_1} 2^{\hat{a}_2} 3_{\hat{b}_3}4_{\hat{b}_4}]  \over s_{12} \, s_{23}\, s_{31} } +{\rm sym} \, . \nonumber \\
\eea
We symmetrize $\hat{a}_3, \hat{b}_3$ and $\hat{a}_4, \hat{b}_4$ for the graviton multiplets, and $[1_{\hat{a}_1} 2_{ \hat{a}_2} 3_{ \hat{a}_3} 4_{ \hat{a}_4}] = \epsilon_{ABCD} \tilde{\lambda}^A_{1\hat{a}_1} \tilde{\lambda}^B_{2\hat{a}_2} \tilde{\lambda}^C_{3\hat{a}_3} \tilde{\lambda}^D_{4\hat{a}_4}$, and $\delta^8 (Q)=\delta^8(\sum_{i=1}^4 \lambda_{i,a}^A \eta^{Ia}_{i})$.

\vspace{-0.15cm}

\section{The K3 Moduli Space from Soft Limits}

Type IIB string theory compactified on $\mathrm{K}3$ has a well studied moduli space described by the coset~\cite{Aspinwall:1996mn},
\bea
\!\!\!\! \mathcal{M}_{(2,0)} = SO(5,21;\mathbb{Z})\backslash SO(5,21) / (SO(5) \times SO(21)). 
\eea
The discrete group is invisible in the supergravity approximation, so we concern ourselves with the local form of the moduli space of supergravity theory, namely ${SO(5, 21) \over {SO(5) \times SO(21)}}$. It has dimension of $105$, which corresponds precisely to the $105$ scalars in the $21$ tensor multiplets. These scalars are Goldstone bosons of the breaking of $SO(5, 21)$ to $SO(5) \times SO(21)$, which are the R-symmetry and flavor symmetry, respectively. Therefore, the scalars obey soft theorems, which are the
tools to explore the structure of the moduli space directly from the S matrix~\cite{ArkaniHamed:2008gz}. 

We find that the amplitudes behave like pion amplitudes with ``Adler's zero"~\cite{Adler:1964um} in the single soft limit. Indeed for ${p_1 \rightarrow 0}$, we find,
\bea
A_n(\phi_1^{f_1}, 2, \ldots, n) \rightarrow \mathcal{O}(p_1) \, ,
\eea
and the same for other scalars in the tensor multiplets. The commutator algebra of the coset space may be explored by considering double soft limits for scalars. Begin with the flavor symmetry, we find for $p_1, p_2 \rightarrow 0$ simultaneously
\bea
\!\!\!\! A_n(\phi^{f_1}_1, \bar{\phi}^{f_2}_2, \ldots) \rightarrow 
{1\over 2}\sum_{i=3}^n { p_i\cdot (p_1 - p_2) \over p_i\cdot (p_1 + p_2)}  R^{f_1\, f_2}_i A_{n-2}  \, ,
\eea
where $f_i$'s are flavor indices, and $R^{f_1\, f_2}_i$ is a generator of the unbroken $SO(21)$, which may be viewed as the result of the commutator of two broken generators. $R^{f_1\, f_2}_i$ acts on superfields as 
\bea
R^{f_1\, f_2}_i \Phi_i^{f_2} &=&\Phi_i^{f_1}, \quad R^{f_1\, f_2}_i \Phi_i^{f_1} =- \Phi_i^{f_2},
\cr 
R^{f_1\, f_2}_i \Phi_i^{f_3} &=&0, \quad \quad R^{f_1\, f_2}_i \Phi_{i,\hat{a} \hat{b}} =0,
\eea 
where $f_3 \neq f_1, f_2$.
Therefore, the generator exchanges tensor multiplets of flavor $f_1$ with ones of $f_2$, and sends all others and the graviton multiplet to $0$. 

To study the $SO(5)$ R-symmetry generators we take soft limits of two scalars which do not form a R-symmetry singlet. For instance 
\bea
\!\! A_n(\bar{\phi}_1, {\phi}^{IJ}_2, \ldots)  \rightarrow
{1\over 2}\sum_{i=3}^n { p_i\cdot (p_1 - p_2) \over p_i\cdot (p_1 + p_2)} R^{IJ}_{i}  A_{n-2}   \, ,
\eea
with $R^{ IJ}_{i} ={  \eta_{i, a}^I  \, \eta_{i}^{J, a} }$.
Similarly, other choices of soft scalars lead to the remaining R-symmetry generators: 
\bea
R_{i, IJ} ={  \frac{\partial}{\partial \eta_{i, a}^I }
\frac{\partial}{\partial \eta_{i}^{J, a}} }, \quad  R^{ I}_{i, J} =  \eta_{i, a}^I  { \partial \over \partial  \eta_{i, a}^{\, J}}.
\eea
Finally, we consider the cases where soft scalars carry different flavors and do not form an R-symmetry singlet. This actually leads to new soft theorems:
\bea \boxed{
\!\! A_n(\bar{{\phi}}^{f_1}_1, {\phi}^{f_2, IJ}_2, \ldots)  \rightarrow
\sum_{i=3}^n { p_1\cdot p_2 \over p_i \cdot (p_1 {+} p_2)}  R^{f_1 f_2}_i R^{IJ}_{i} A_{n-2} }   \hspace{2mm}
\eea
and similarly for other R-symmetry generators. The results of the soft limits now contain both flavor and R-symmetry generators, reflecting the direct product structure in ${SO(5, 21) \over {SO(5) \times SO(21)}}$. This is a new phenomenon that is not present in pure $(2,0)$ supergravity~\cite{ArkaniHamed:2008gz, Chen:2014cuc}. 

The above soft theorems may be obtained by analyzing how the integrand and the scattering equations behave in the limits. For instance, the vanishing of the amplitudes in the single-soft limits is due to
\bea 
\int d\mu_n^{6D} \sim \mathcal{O}(p_1^{-1}), \quad {\rm det}^{\prime} S_n \sim \mathcal{O}(p_1^2),
\eea
and the rest remains finite. The double-soft theorems require more careful analysis along the lines of, e.g.~\cite{Volovich:2015yoa}. The structures of double-soft theorems, however, are already indicated by knowing the four-point amplitudes given in (\ref{eq:4pts}), since important contributions are diagrams with a four-point amplitude on one side such that the propagator becomes singular in the limit~\cite{comment:3ptsoft}. Finally, we have also checked the soft theorems explicitly using our formula (\ref{eq:multi-tensor}) for various examples.   

\vspace{-0.15cm}
\section{4D $\mathcal{N}=4$ Einstein-Maxwell Theory}

One can dimensionally reduce 6D $(2,0)$ supergravity to obtain 4D $\mathcal{N}=4$ Einstein-Maxwell theory. The tree-level amplitudes of this theory capture the leading low-energy behavior of Type IIB (or Type IIA) superstring theory on $\mathrm{K}3\times T^2$. 

The reduction to 4D can be obtained by decomposing the 6D spinor as $A \rightarrow \alpha = 1,2, \,\, \dot{\alpha} =3,4$. The compact momenta are $P_i^{\alpha \beta} = P_i^{\dot{\alpha}\dot{\beta}}=0$; this is implemented by 
$ \lambda^A_a \rightarrow \lambda^{\alpha}_+ = 0$ and $\lambda^{\dot \alpha}_- = 0.$

The 6D tensor superfield becomes an $\mathcal{N}=4$ vector multiplet in 4D, in a non-chiral form~\cite{Huang:2011um, Heydeman:2017yww}, 
\begin{align}
\Phi(\eta_a)& \rightarrow  V_{\mathcal{N}=4}(\eta_+, \eta_-) =  \phi  +  \eta^{\hat I}_-  \psi^-_{\hat I} 
+  \dots \nonumber \\
+& \, (\eta_+)^2  A^{+} 
+ \, (\eta_-)^2 A^{-} +\dots
+  (\eta_+)^2  (\eta_-)^2  \bar{\phi} \, .
\end{align}
Dimensional reduction of $\Phi^{\hat{a} \hat{b}}(\eta)$ is analogous. It separates into $3$ cases, where $\Phi^{\hat{+} \hat{-}} \rightarrow V_{\mathcal{N}=4}(\eta_+, \eta_-)$, and $\Phi^{\hat{+} \hat{+}}, \, \Phi^{\hat{-} \hat{-}}$ become a pair of positive and negative-helicity graviton multiplets
\begin{align} \label{eq:4D_superfield}
&\Phi^{\hat{+} \hat{+}}(\eta_a) \rightarrow \mathcal{G}^+_{\mathcal{N}=4}(\eta_+, \eta_-) =  A^+  +  \eta^{\hat I}_-  \psi^{-+}_{\hat I} 
+   \dots \nonumber \\
+&\,  (\eta_+)^2  G^{++} +  (\eta_-)^2 \phi
+  \dots
+  (\eta_+)^2  (\eta_-)^2  \bar{A}^+ \, ,\\
&\Phi^{\hat{-} \hat{-}}(\eta_a) \rightarrow \mathcal{G}^-_{\mathcal{N}=4}(\eta_+, \eta_-) =  \bar{A}^-  +  \eta^{\hat I}_-  \Psi^{--}_{\hat I} 
+   \dots \nonumber \\
+& \, (\eta_-)^2  G^{--} +  (\eta_+)^2 \bar{\phi} 
+\dots
+  (\eta_+)^2  (\eta_-)^2  {A}^- \, .
\end{align}
We see the on-shell spectrum of the 4D supergravity theory consists of the $\mathcal{G}^+$ and $\mathcal{G}^-$ superfields coupled to $22$ $\mathcal{N}=4$ Maxwell multiplets. 

We are now ready to perform the dimensional reduction on (\ref{eq:multi-tensor})~\cite{comment0}. First, the 6D measure reduces to
\bea
d\mu^{\rm 4D} = {\prod_{i=1}^n d\sigma_i \prod_{k=0}^d d^{2}\rho_k \prod_{k=0}^{\tilde{d}} d^{2} \tilde{\rho}_k \over \vol(\slc_{\sigma} \times \glc)} {1 \over R(\rho) R(\tilde{\rho})} \prod_{i=1}^n E_i^{4D} \nonumber 
\eea
where $R(\rho)$, $R(\tilde{\rho})$ are the resultants of the polynomials 
\bea
\rho^{\alpha}(\sigma)=\sum^d_{k=0}\rho^{\alpha}_k\, \sigma^k , \quad
\tilde{\rho}^{\dot \alpha}(\sigma)=\sum^{\tilde{d}}_{k=0}\tilde{\rho}^{\dot \alpha}_k\, \sigma^k ,
\eea
with $d+\tilde{d}=n-2$, and the 4D scattering equations are given by
\bea
E_i^{4D} = \delta^4\left( p^{\alpha {\dot \alpha}}_i - {\rho^{\alpha}(\sigma_i) \tilde{\rho}^{\dot \alpha}(\sigma_i) \over \prod_{j \neq i} \sigma_{ij} } \right).
\eea
The $2\times 2$ matrix $(\tilde{W}_i)_{ab}$ reduces to
\bea
\!\!\! (\tilde{W}_i)_{\hat{+}\hat{+}} = (\tilde{W}_i)_{\hat{-}\hat{-}} =0, \,\,\, (\tilde{W}_i)_{\hat{+}\hat{-}} =t_i , \,\,\,  (\tilde{W}_i)_{\hat{-}\hat{+}} = \tilde{t}_i .
\eea 
with
$
t_i = { \lambda^{\alpha}_i \over  \rho^{\alpha}(\sigma_i) } ,\,
\tilde{t}_i = {\tilde{ \lambda}^{\dot \alpha}_i \over  \tilde{\rho}^{\dot \alpha}(\sigma_i) }
$
(independent of $\alpha, {\dot \alpha}$), and $t_i \tilde{t}_i = \prod_{j\neq i} {1\over \sigma_{ij}}$. As for the integrand, the parts that reduce to 4D non-trivially are
\bea \label{eq:reduction}
\tilde{M}^{n_1}_{\hat{a} \hat{b}}  \rightarrow 
 \tilde{T}^{n_1}_{\hat{a} \hat{b}}, \quad \, \,
 {\rm det}^{\prime} S_n \rightarrow  R^2(\rho) R^2(\tilde{\rho}) {V}^{-2}_{n}.
\eea 
Assume we have $m_1$ $\mathcal{G}^+$ superparticles and $m_2$ $\mathcal{G}^-$, with $m_1 + m_2 = n_1$~\cite{comment1}, we find $\tilde{T}^{n_1}$ is given by 
\bea
\tilde{T}^{n_1} = T_+^{m_1} \,  T_-^{m_2}= \left(V^2_{m_1} \prod_{i \in m_1} t^2_i \right) \left( V^2_{m_2}  \prod_{j \in m_2}\tilde{t}^2_j \right),
\eea 
where $V_{m_1}=\prod_{i<j}\sigma_{ij}$ for $i, j\in m_1$, and similarly for $V_{m_2}$. 
We therefore obtain a general formula for the amplitudes of 4D $\mathcal{N}=4$ Einstein-Maxwell theory:
\bea \label{eq:4D-general}
\!\!\! A^{\mathcal{N}=4}_n =\! \int \! d\mu^{\rm 4D}  \, { {\rm Pf} \mathcal{X}_{n_2} \over V_{n_2} V_n} \, {T}^{m_1}_+ \, {T}^{m_2}_-  
 R^2(\rho) R^2(\tilde{\rho})\, \Omega^{\mathcal{N}=4}_F , 
\eea 
where $\Omega^{\mathcal{N}=4}_F$ implements the 4D $\mathcal{N}=4$ supersymmetry, arising as the reduction of $\Omega^{(2,0)}_F$,
\bea 
\!\!\! \Omega^{\mathcal{N}=4}_F= \prod_{k=0}^d \delta^2(\sum_{i=1}^n t_i \sigma_i^k \eta_{i+}^I)\prod_{k=0}^{\tilde{d}} \delta^2(\sum_{i=1}^n \tilde{t}_i \sigma_i^k \eta_{i-}^{\hat I}).
\eea 
The formula should be understood as summing over $d,\tilde{d}$ obeying $d+\tilde{d}=n-2$. However, it is clear from the superfields that we should require
\bea
d={n_2\over 2} + m_1 - 1, \quad \quad 
\tilde{d}= { n_2\over 2} + m_2 - 1,
\eea
recall $n_2$ is even. Therefore, for a given number of photon and graviton multiplets, the summation over sectors becomes a sum over different $m_1, m_2$. 
We have checked (\ref{eq:4D-general}) against many explicit amplitudes, and also verified that the integrand is identical to that of~\cite{Cachazo:2014xea} for certain component amplitudes. 

\vspace{-0.15cm}

\section{Discussion and Conclusion}
We have presented a formula for the tree-level S matrix of 6D $(2,0)$ supergravity. The formula for single-flavor tensor multiplets is constructed via a SUSY reduction of the one for $(2,2)$ supergravity. We observed important simplifications in deriving the formula, particularly the appearance of the object ${\rm Pf}X_n$, crucially for the generalization to $21$ flavors required for $(2,0)$ supergravity. By studying soft limits of the formula, we were able to explore the moduli space of the theory. Via dimensional reduction, we also deduced a new formula for amplitudes of 4D $\mathcal{N}=4$ Einstein-Maxwell. Since 6D $(2,0)$ supergravity has a UV completion as a string theory, it would be of interest to extend our formula to include $\alpha'$ corrections, perhaps along the lines of~\cite{Mizera:2017sen}. Also, a recent paper~\cite{Geyer:2018xgb} introduces an alternative form of the scattering equations that treats even and odd points equally, but uses a different formalism for supersymmetry. It will be interesting to study our formula into this formalism. 

Our results provide an S matrix confirmation of various properties of $(2,0)$ supergravity and the dimensionally reduced theory as predicted by string dualities. While the 10D theory has a dilaton that sets the coupling, in 6D this scalar is one of the $105$ moduli fields, and appears equally with the other $104$ scalars. If one considers the compactification on $\text{K3} \times T^2$, standard U-dualities imply equivalence to the Type IIA superstring theory on the same geometry or the heterotic string theory compactified to 4D on a torus. The formulas discussed in this article apply to all these cases, at least at generic points of the moduli space.

\section{Acknowledgements}
We thank Nima Arkani-Hamed, Yvonne Geyer and Shu-Heng Shao for very helpful discussions. We also thank Freddy Cachazo, Alfredo Guevara, and Sebastian Mizera for discussions and correspondence on related topics. C.W. is supported by a Royal Society University Research Fellowship No. UF160350. S.Q.Z. is supported by the Royal Society grant RGF\textbackslash R1\textbackslash 180037. M.H.\ would like to thank S.S.~Gubser and Princeton University for their hospitality, and work done at Princeton was supported in part by the Department of Energy under Grant No.~DE-FG02-91ER40671, and by the Simons Foundation, Grant 511167 (SSG).
M.H. and J.H.S. are supported in part by the Walter Burke Institute for Theoretical Physics at Caltech and by U.S. DOE Grant DE-SC0011632.

\begin{thebibliography}{99}
\bibitem{comment-moduli}
  {The string-theory moduli space has singularities
at fixed points of its $SO(5,21;\mathbb{Z})$ duality group. At such points one or more tensor multiplets are replaced by non-Lagrangian $(2,0)$ CFTs, and a perturbative analysis is no longer possible. Therefore, the amplitudes presented in this article are applicable at generic points in the moduli space where we can treat the tensor multiplets as abelian.}
\bibitem{comment-susyred}
  {For the superfield of the tensor, $\Phi(\eta)$, one may choose different $I, J$; however, only the case of $I=1, J=2$ leads to non-vanishing results when we integrate away $\tilde{\eta}$'s from the amplitudes of $(2,2)$ supergravity.}
\bibitem{comment-simplification}
  {The identity may be understood by studying the zeros and singularities on both sides of the equation, we have also checked it explicitly up to $n=12$.}
\bibitem{comment:even-odd}
  It should be emphasized that the artificial difference between the formulas of even- and odd-point amplitudes is due to the peculiar property of 6D scattering equations in the rational map form, not specific to the chiral $(2,0)$ supergravity we consider in this article.
\bibitem{comment:3ptsoft}{For the study of the moduli space, the scalars we choose do not form a singlet, therefore the most singular diagrams with a three-point amplitude on one side are not allowed. If instead the two scalars do form a singlet, then such diagrams dominate and contain a propagating soft graviton. The soft theorem for such a pair of scalars is simply given by Weinberg's soft graviton theorem attached with the three-point amplitude, which we have also checked using our formula.}
\bibitem{comment0} 
  {As discussed in~\cite{Cachazo:2018hqa}, the reduction to 4D kinematics may be subtle in this formalism that only the middle sector of even $n$ (or next to the middle sector of odd $n$) is straightforward. The strategy we adapt here is that once we obtain the formula for the middle (or next to the middle) sector, we then generalize them for other sectors, which is straightforward.}
\bibitem{comment1} 
  {We do not consider $\Phi^{\hat{+} \hat{-}}$ here since they are identical to the vector, for which we have already included, as shown in (\ref{eq:4D_superfield})}.
\end{thebibliography}

\end{document}